# Hanbury Brown – Twiss type correlations with surface plasmon light

Sándor Varró*, Norbert Kroó, Dániel Oszetzky, Attila Nagy and Aladár Czitrovszky

*Research Institute for Solid State Physics and Optics of the Hungarian Academy of Sciences*, *Budapest, Hungary*

*Corresponding author. Email: varro.sandor@wigner.mta.hu, varro@mail.kfki.hu
 Postal address: H-1525 Budapest, POBox 49, Hungary

# Hanbury Brown and Twiss type correlations with surface Plasmon light

Sándor Varró, Norbert Kroó, Dániel Oszetzky, Attila Nagy and Aladár Czitrovszky

**Abstract**. Intensity-intensity correlations are studied for light signals stemming from the spontaneous decay of surface plasmon oscillations, generated in the Kretschmann geometry. Non-classical photon statistics and the transition from antibunching to bunching of the electron counts have been found experimentally and analysed on a new theoretical basis.

## 1. Introduction

Surface plasmon physics still raises many fundamental questions concerning radiation-matter interaction and optics [1-8], which are interesting in themselves. At the same time, this rapidly growing field of research will surely impact many branches of science and technology [9]. Recently there has been a renoved interest, in general, in the investigation of higher-order coherence properties and correlations of Hanbury Brown and Twiss type between both bosons [10-11] and fermions [12], and there are various questions and interpretations on the non-classical nature of such correlations. Concerning surface plasmon oscillations (SPOs), studies devoted to explore excitations in the single-photon regime [13] and transverse coherence properties [14] has been published. In our recent research [15-17] on the physical properties of the SPOs themselves, and on light emitted by them, we have already found indications for the possible sub-poissonian nature of the emission which has been probed in photon counting experiments. Initiated by these observations with surface plasmon excitations, we have developed a new experimental setup for the measurement of the additional statistical properties of the light generated by SPOs.

Non-classical effects in light-matter interactions have long been studied in atomic physics [18-19], and in the meantime they have become standard subjects of textbooks on quantum optics [20-23]. Concerning surface plasmons, both the conversion of the incoming photons to SPOs [13] and the decay process [15], or simply the interaction of photons with the free electrons [24-28] can cause non-classical effects, like squeezing (owing to the large density of electrons in a metal). The main challenge, for both the experimenter and for the theorist, in studying the coherence properties of surface plasmon light is due to the smallness of the decay time (on the order of 100 femtoseconds), in comparison with the sampling time. In quantum optics, in the early days of photon counting experiments, such questions have already been thoroughly discussed [29]. When one uses a data acquisition system with 12.5 ns time resolution, (to which, on top, the dead time of about 60 ns of the detectors is superimposed, as was the case in our experiments), and measures some systematic deviation from the shot noise level, one has to clearly separate possible instrumental artifacts in the course of data processing. In Section 2, in the theoretical part of the present paper, we shall consider these kind of questions on the basis of a new mathematical description of photon counting measurements, adapted to the situation under discussion. We shall prove that, in spite of the enormous (six order of magnitudes) factor between the effective time resolution and the characteristic time of the plasmon decay, it is possible to isolate non-trivial effects in the experimental data. In Section 3 we shall describe our experimental setup, and present few illustrative results concerning the second order correlations between the delayed numbers of photoelectron counts. In Secion 4 a brief summery closes our paper.

**2. Theoretical considerations on coherence properties in surface plasmon decay**

In the present section we briefly summarize some basic properties of the surface plasmon oscillations (SPO), which may be relevant for the interpretation of our results concerning correlation measurements on light signals emitted by the SPOs. Regardless of how the SPO has been generated, when it spontaneously decays, its life-time has an exponential distribution. Since the typical count rate in our experiments was 200000/s and the plasmon decay time is ~ $10^{-13}$s, there is a negligible probability for more than one photon emission from the sample during the effective detection time ~ 75ns. This means that, in fact, here we are dealing with a non-classical single-photon source. In the second part of this section we present a new analytic approach, based on conventional probability theory, in order to determine the photon counting statistics along this physical picture, and calculate second-order correlations, too.

*Classical kinematics and dynamics of surface plasmon generation and decay*

In the case of the Kretschmann geometry [30-32] considered throughout the present paper, the complex magnetic induction and the electric field strength of a p-polarized (transverse magnetic; TM) plane wave can be expressed in a condensed form as

$$\{\boldsymbol{B}; \boldsymbol{E}\} = \{\boldsymbol{u}_y B(z); -(c/\omega\varepsilon)[i\boldsymbol{u}_x \partial_z B(z) + \boldsymbol{u}_z k_x B(z)]\} \exp[i(k_x x - \omega t)]. \quad (1)$$

Here $\boldsymbol{u}_{x,y,z}$ denote unit vectors pointing along the corresponding Cartesian coordinate axes, whose origin is at the middle of the sample at the $z = 0$ plane (which defines the metal-vacuum interface). The metal-dielectric interface is defined by the plane $z = d$, where, for instance in our experiments the thickness of the gold layer was $d = 45nm$, i.e. about two times larger than the skin depth at optical frequencies. The function $B(z)$ usually has a character of plane waves $\exp[\pm ik_z z]$ (either propagating or evanescent), and (owing to the continuity of $E_x$ and $B_y$ at the interfaces) both $B(z)$ and $\partial_z B(z)$ have to be continuous. By using the experimental data of Johnson and Christy [33], the permittivity of the gold layer (called here region 2: $0 < z < d$) can be taken $\varepsilon_2(\omega_0) = -\varepsilon_R + i\varepsilon_I \approx -25.82 + i1.63$ at the central frequency of the incoming light. For a glass substrate (region 1: $z > d$) and air (region 3: $z < 0$) we have $\varepsilon_1 = 2.25$ and $\varepsilon_3 = 1$, respectively. Under such circumstances the calculated angle of total reflection is $\theta_t = 41.8°$, and the attenuated total reflexion takes place at the critical angle $\theta_c \approx 42.8° = \theta_t + 1°$. The plasmon resonance conditions – which define the eigenmodes of the dielectric-metal-vacuum system – can be expressed by the well-known (approximate) implicit dispersion relation $k_{sp} \approx (\omega/c)[\varepsilon_2(\omega)\varepsilon_3/(\varepsilon_2(\omega) + \varepsilon_3)]^{1/2}$. At the metal-vacuum interface the components of the resonant wave vector can be well approximated by the expressions

$$k_{2x} = k_{sp} = \frac{\omega_0}{c}\left[1 + \frac{1}{2(\varepsilon_R-1)} + i\frac{\varepsilon_I}{2(\varepsilon_R-1)^2}\right], \quad k_{2z} = \frac{\omega_0(\varepsilon_R-1)^{1/2}}{c}\left[1 + i\frac{\varepsilon_I}{2(\varepsilon_R-1)}\right]. \quad (2)$$

On the basis of the empirical values $\varepsilon_2(\omega_0) = -\varepsilon_R + i\varepsilon_I \approx -25.82 + i1.63$, Eq. (2) shows that the dispersion is close to the light dispersion in vacuum, and the propagation length $L_{sp} \approx (\varepsilon_R - 1)^2 \lambda_0/2\pi\varepsilon_I$ is much larger than the wavelength of the incoming radiation. For a *He-Ne* laser light ($h\nu_0 = 1.963 eV$, $\lambda_0 = 633 nm$) we have $L_{sp} \approx 38 \mu m \approx 60\lambda_0$. In the metal layer (region 2: $0 < z < d$), along the perpendicular direction to the metal-vacuum interface, the energy density of the SPO field decreases on the scale $\delta_2 = \lambda_0/4\pi(\varepsilon_R - 1)^{1/2} \approx 10 nm \approx \lambda_0/63$. This means that the energy density at the opposite interface $z = d = 45 nm$ drops down to 1% of its value at $z = 0$. Along the other transverse direction the excitation extends over the spot size $L_y \approx 25 \mu m \approx 40\lambda_0$, thus the area $A_{sp} = L_{sp}L_y$ occupied by the SPOs is at least $A_{sp} \approx 2400\lambda_0^2 = 9.6 \times 10^{-6} cm^2$ in the case under discussion. On the basis of our recent theoretical work [17], the relative spectral width of the emitted light is estimated as $\Delta\lambda/\lambda_0 = 0.018$, which in the present case corresponds to a frequency spread $\Delta\nu = 8.526 \times 10^{12} s^{-1}$. In addition we note that, according to our earlier study [35], the relaxation parameter $\mu$ of the light-induced (collective) surface current turns out to be $\mu = 0.032\omega_0$ for a gold target. Thus, the associated spontaneous life-time is expected to be on the order $\tau_s = 1/\mu \sim 100 fs$, which is consistent with the value of the spectral width $\Delta\nu$, presented a few lines before.

*Statistics of counting events in case of large sampling times*

As we have already mentioned in the introduction, in principle, both the conversion of the incoming photons to SPOs [13] and the decay process [15], or simply the interaction of photons with the free electrons [24-28] can cause non-classical effects [19], like squeezing (owing to the large density of the electrons). These effects are usually described by the quantum theory of light [22], which accounts for the discreteness of the occupation of the modes of the electromagnetic radiation. On the other hand, quite much can be said about *quantum* correlations by using classical *discrete* probability distributions [36]. Here we apply a simple approach [37], which, perhaps makes the connection of the mathematical formalism with the actual physical problem easier to follow.

Let us assume that the life time of an excitation in the plasmon field is a random variable with exponential distribution $F(t) = 1 - e^{-\mu t}$ ($t \geq 0$), where $\mu$ is the decay constant [38]. We have expressed the probability $F_n(t)$ of $n$ separate photon emission in the main propagation direction by a weighted $n$-fold convolution,

$$F_n(t) = P(n, \sigma\mu t), \quad P(n, x) \equiv \frac{1}{(n-1)!} \int_0^x du\, u^{n-1} e^{-u} = \frac{\gamma(n,x)}{\Gamma(n)} =. \tag{3}$$

Here $\gamma(\alpha, x) \equiv \int_0^x du\, u^{\alpha-1} e^{-u}$ is the incomplete gamma function [39], and $\sigma$ is the probability that one photon falls in the transverse modes covered in the course of emission. By using the integral representation above, it is an easy matter to show that the sum rule $\sum_{n=1}^{\infty} P(n, \sigma\mu t) = \sigma\mu t$ holds, which is not an unexpected result. The sum of the probabilities of all photon emissions is proportional with the elapsed time. The proportionality factor is the product of the original decay constant and the average weight of one photon emission in a particular direction. Thus, the product $\sigma\mu$ is the average photon flux in that direction.

The life-time of an excited electron in its ground state is characterized by the distribution function $F^e(t) = 1 - e^{-\kappa t}$ ($t \geq 0$), which describes the depletion of the ground state, where $\kappa$ is the excitation (ionization) parameter. The probability of $k$ separate ejection by the total time $t$ is given by the distribution $F_k^e(t) = P(k, \kappa t) = \gamma(k, \kappa t)/\Gamma(k)$, where $\gamma$ denotes the incomplete gamma function. The total probability of that all the activated electrons are ejected is again proportional with the elapsed time, $\sum_{k=1}^{\infty} P(k, \kappa t) = \kappa t$. The distribution of the *random number $v_\tau$ of the electron ejection in a fixed interval* $(0, \tau)$ can be determined in the same manner we have seen for the random photon emission, and we again encounter with the Poisson distribution; $P^e(v_\tau = k) = F_k^e(\tau) - F_{k+1}^e(\tau) = \frac{\lambda^k}{k!} e^{-\lambda}$, where $\lambda = \kappa\tau$. Of course, the statistics of the counts must depend on the statistics of the photon arrivals at the detector. The Poisson distribution alone merely represent a backround noise, the shot-noise. The most straightforward way to take into account the nontrivial influence of the light's statistics, is that we weight the Poisson distribution of the electrons with the photon distribution, term by term. This procedure is particularly justified in the present case of a low-intensity emitter. Within this approximation, the counting events are modelled by the set of weights $\{W_0, W_1, \ldots, W_n, \ldots\}$ of the type

$$W_n = W_0 s \frac{\lambda^n}{n!} P(n, \sigma\mu t) \quad (n \geq 1), \tag{4a}$$

$$W_0 = \left[1 + s \sum_{n=1}^{\infty} \frac{\lambda^n}{n!} P(n, \sigma\mu t)\right]^{-1} = \left[1 + s\sqrt{\lambda} \int_0^{\sigma\mu t} \frac{du}{\sqrt{u}} I_1(2\sqrt{\lambda u}) e^{-u}\right]^{-1}, \tag{4b}$$

where the first equation in Eq. (4a) comes from the normalization condition $\sum_{n=0}^{\infty} W_n = 1$. On the basis of Eq. (3), we have expressed the infinite sum by an integral, where we have also used the power series representation of the modified Bessel function $I_1(z)$ of first kind of order 1[39]. The parameter $s$ introduced in the statistical weights for $n \geq 1$ in Eq. (4a) accounts for an overall efficiency of coupling the radiation to the beam splitter. The *Ansatz* introduced in Eq. (4a), corresponds to the intuitive picture that each photon deliberates exactly one electron, owing to the conservation of energy. At this point we would like to note, that the gamma distribution in Eq. (3) can also be obtained formally from Mandel's formula (see eg. Ref. [36], formula (9.2-1)), by assuming a chaotic continuous energy distribution stemming from the plasmon. In our description the chaoticity comes from the possibility of distributing the emitted single photons among the very large number of spatial transverse modes. Mathematically, the essence of the ansatz used above in Eq. (4a) lies in that we do not consider more general terms of the type $\frac{\lambda^n}{n!}P(m, \sigma\mu t)$ with different $n$ and $m$ indeces. From such "direct products" many other types of distributions could be compiled by using finite or infinite weighted sums, for instance. In the present case we think that, the inclusion of only the "diagonal terms" $\frac{\lambda^n}{n!}P(n, \sigma\mu t)$ is perfectly justified, because the source produces a "very dilute single-photon beam"; i.e. one detected electron must strictly be associated only to one emitted photon.

The quantities of primary interest in the present paper are the correlations of the counting events in the two detectors, which are quantified by the following expressions

$$K \equiv \overline{N_1 N_2}/\overline{N}_1\,\overline{N}_2 \quad , \quad R \equiv \overline{\Delta N_1 \Delta N_2}/\overline{N}_1\overline{N}_2 = K - 1 = \frac{1}{\langle n \rangle}(F - 1), \qquad (5a)$$

$$F \equiv \frac{\langle \Delta n^2 \rangle}{\langle n \rangle}, \qquad \langle n \rangle \equiv \sum_{n=0}^{\infty} n W_n, \qquad \langle \Delta n^2 \rangle \equiv \langle (n - \langle n \rangle)^2 \rangle. \qquad (5b)$$

In Eq. (5b) we have introduced the Fano factor $F$, which is, in general, an important characteristic of *any* distribution. In quantum optics, the well-known $Q$ parameter introduced by Mandel, is a special case of $\tilde{Q} \equiv F - 1$ for photon distributions. In Eqs. (8a-b) we have explicitly written out the definitions of the two type of mean values, namely the expectation values concerning the counting events (this is denoted by upper dash), and the ensemble averages over the source distributions (which are denoted by angular brackets). The moments of the distribution $\{W_n\}$, defined in Eq. (4a), can be conveniently calculated by introducing the generating function, which reads

$$G(z) \equiv \sum_{n=0}^{\infty} W_n z^n = W_0 \left[1 + 2s\sqrt{\lambda z \sigma \mu t} \int_0^1 du I_1(2\sqrt{\lambda z \sigma \mu t}\,u) e^{-\sigma \mu t u^2}\right], \qquad (5)$$

where we have used the same procedure as that leading to Eq. (4b), and introduced a new integration variable (denoted by the same letter $u$). As is well-known, for istance, the first moment of the distribution $\{W_n\}$ can be calculated from the first derivative of the generating function; $\langle n \rangle = [\partial G(z)/\partial z]_{z=1}$, etc. From the functional form of the generating function, displayed in Eq. (5), it is easily seen that all the moments have to depent, on one hand, on the pump parameter $s$, and on the other hand on the product $\lambda\sigma\mu t$, where $t$ is time of the joint excitation of the detectors.

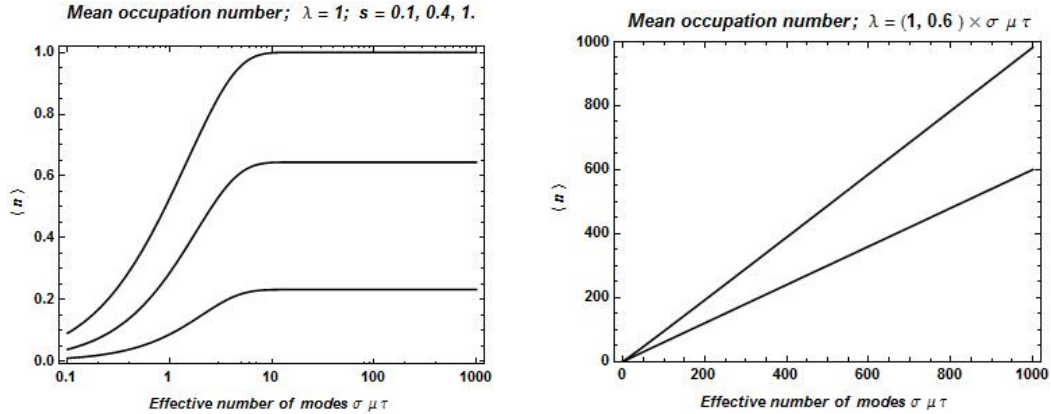

**Figure 1.** Left (a): shows in a log-linear plot the dependence of the mean occupation number on the dimensionless interaction time $\sigma\mu\tau$, for three values of the coupling parameter $s$. Here the parameter $\lambda$ has been kept fixed. Right (b): same as in (a), but now we have taken $\lambda$ being proportional with interaction time: $\lambda = \eta \times \sigma\mu\tau$. In this case $\eta$ can be considered as a quantum efficiency (for which we have taken the values 1 and 0.6. Since the figure shows exact linear dependence, the mean number of electron counts is simply proportional with $\langle n \rangle$ in this case. This proportionality is not affected by the value of the coupling parameter $s$.

In Fig. 1 the mean occupation number of the distribution $\{W_n\}$ is shown in two physically different cases; on the left the parameter $\lambda$ has been kept fixed, on the right it has been taken proportional with the effective number of modes $\sigma\mu\tau$ (which is the number of relevant modes in the counting process). This latter figure completely corresponds to the intuitive expectation that the mean number of true counts $\lambda = \kappa\tau = \eta \times \sigma\mu\tau$ should be proportional with the mean number of absorbed photons, because the detection process is linear. If we look at the complicated form of the distribution in Eqs. (4a-b), and the generating function in Eq. (5), then this is far not a trivial mathematical result. The other remarkable result is the found independence on the coupling parameter $s$. This parameter quantifies the pumping of the measuring apparatus from the outer source, and it is not a property of the detectors. In the dimensionless variable $\sigma\mu\tau$, the second factor $\mu\tau$ is the ratio of the true (joint) interaction time with the detectors and the coherence time $\tau_c = 1/2\mu$ of the radiation emitted by one electron in the surface plasmon. Thus $2\mu\tau = \tau/\tau_c$ is the number of relevant longitudinal modes. The effective interaction time during one elementary detection $\tau = 12.5 ns + 63.5 ns$ is the sum of the nominal resolution and the dead time $> 60 ns$. The size of the first factor $\sigma$ can roughly be estimated by using the Van Cittert – Zernike theorem [36], which states that the radiation coming from a diaphragm with aperture $d$, and reaching another aperture $D$ at distance $z$, is transversally coherent if in the expression $2\pi D d/z\lambda_0 = (2/\pi)M_{tr}$ the number $M_{tr}$ is not larger than 2. If the radiation stems from a center behind the aperture $d$ at distance $z_1 (\ll z)$, then $M_{tr} \approx \pi^2 d^2/z_1\lambda_0 = \pi A/z_1\lambda_0$ (because $D/z \approx d/z_1$), where $A = \pi d^2$ is the area of the first aperture. If $z_1 = \lambda_0$ then $M_{tr} = \pi A/\lambda_0^2$ is von Laue's expression for the total number of transverse modes (degrees of freedom) in the half space $(z > 0)$. According to this formula, for a radiation center situated in the skin depth at distance $z_1 = \delta_2 = \lambda_0/63$ from the surface, and radiating through the plasmon area $A_{sp} \approx 2400\lambda_0^2$ (which serves as an aperture), we estimate $M_{tr} = 152000$. Such a single-photon source has an extremely low transverse degeneracy $1/M_{tr} < 10^{-5}$, and the parameter $\sigma$ is very small.

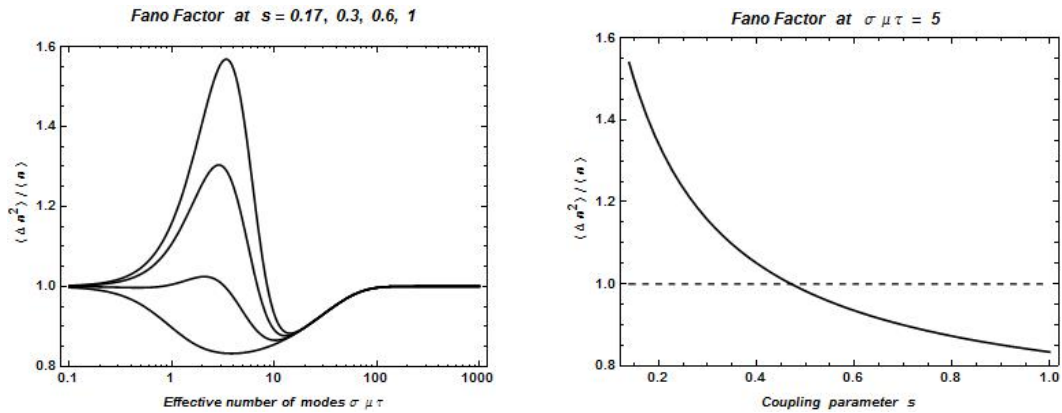

**Figure 2.** Left (a): shows in a log-linear plot the Fano factor $F = \langle\Delta n^2\rangle/\langle n\rangle$ on the dimensionless interaction time $\sigma\mu\tau$, for four values of the coupling parameter $s$. The uppermost curve corresponds to the lowest value $s = 0.17$, and lowermost one to the value $s = 1$. Right (b): the transition from bunching (super poissonian statistics) to antibunching

(subpoissonian statistics) is illustrated at a particular value of the dimensionless combination $\sigma\mu\tau = 5$. This curve indicate that for very small coupling parameters, the correlation corresponds to more and more disordered (chaotic) distributions, formally $F \to \infty$ as $s \to 0$.

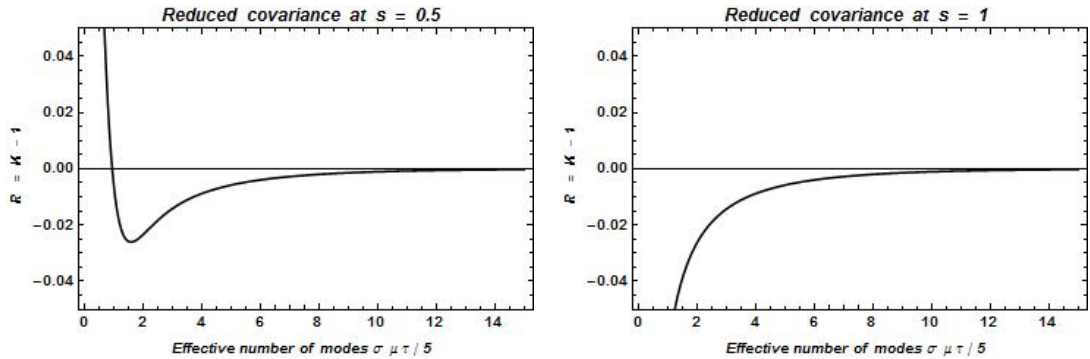

**Figure 3.** Shows the reduced covariance $R = K - 1 = \overline{N_1 N_2}/\overline{N_1}\,\overline{N_2} - 1$ calculated on the basis of Eqs. (5.a-b). This quantity, like the Fano factor plotted in Figs. 2a-b, is very sensitive to the value of the coupling constant. On the left figure we see a transition from anti-bunching to bunching as the scaled delay $\sigma\mu\tau$ decreases. On the right figure, since the coupling constant is already large enough ( $s = 1$ ), no such transition appears, thus the anti-correlation characteristic in the single-photon regime dominates throughout.

It is of interest to know how these theoretical figures compare with our experimental results. More precisely, the question is how can we interrelate the dimensionless quantity $\sigma\mu\tau$ with the experimental parameters? Taking the estimates obtain in the previous sub-section, namely the transverse mode degeracy $\sigma \sim 1/152000$ and $\mu \sim 1/10^{-13}$ s, then $\sigma\mu\tau/5 \sim 1.2 \times (\tau/100ns)$, thus the unit step in Figs. 3a-b should be ment $100ns$, rather than $1\mu s$, as is in the experimental figures in the next section (see Figs.5a-b). The value $\sigma\mu\tau = 5$, used in Fig. 2(b) corresponds to $\sigma = 1/M_{tr} = 1/152000$ (which is transverse mode degeneracy, as has been calculated on page 10), $\mu = 1/10^{-4}ns = 1/100fs$ (which is the plasmon decay rate) and $\tau = 76ns$ (joint interaction time, or effective resolution time). This latter parameter comes from that the data acquisition system used by us has a time resolution 12.5ns, and a dead time of about 63.5ns, thus the effective resolution can be estimated between 72-80ns. That is why, we have chosen $\tau = 76ns$. Thus, according to our experimental conditions, the value $\sigma\mu\tau = 5$ used in plotting Fig. 2 (b) seems to be realistic.

## 3. Experimental results

Based on our previous experiences with surface plasmon excitations [15-17], we have developed a new experimental setup for the measurement of the statistical properties of the light generated by surface plasmon oscillations. A beam of the He-Ne laser was used for the excitation of surface plasmons in a 45*nm* thick gold layer. For the optical setup we used the so-called Kretschmann geometry. The surface plasmons were generated in a thin gold layer by focusing the laser beam of power ~5 mW to the gold layer surface using an F= 40 mm and D = 35 mm lens. To study the statistical properties of photons emitted by surface plasmons we built two measurement setups. With the first one we studied the correlation between the incident (reflected from the gold layer) and the emitted light generated by plasmons, and with the second one the device-independent photon statistics of the emitted light has been studied.

In the first setup (see Fig. 4a.) the beam reflected from the layer and the light emitted from the opposite side of the layer by surface plasmons where measured with Perkin Elmer SPCM - AQR-14 type photon counting modules having high quantum efficiency (~ 60 %) and low noise (dark counts < 100 $sec^{-1}$). The main benefit of the present setup is the possibility to measure the auto-correlation and also the cross-correlation function of the beams. The auto- and cross-correlation function has been measured using a National Instruments NI- 6602 Data Acquisition System having a time resolution 12,5 nsec. The detectable photon rate is determined by the installed FIFO, which enables the measurement of rates up to 200 000 pulse/sec. In the second setup (Fig. 4b.) the cross-correlation function of the light emitted by surface plasmons is measured by splitting the beam by a 50% beam splitter. In this arrangement the fluctuations of the exciting laser and the artifacts of the detectors was eliminated.

In both cases the polarization rotator helps to found the proper position of the plasmon resonance. The whole setup was optically isolated and electronically shielded for reduction of the background noise. The diameter of the focused laser beam was ~ 25 micron, , and the intensity was controlled by an attenuator, so the photon ratecould be varied between 20 000 and 200 000 counts/sec. In case of an s-polarized incoming light, there is still some background signal present in the counts, but this has given count rates always ~ 40 times smaller than those stemming from the excitation with p-polarized light. The whole sampling time, 20 minutes for each count rate and position, was splitted into ten 2-minutes runs to calculate statistical data, like standard deviation and mean values.

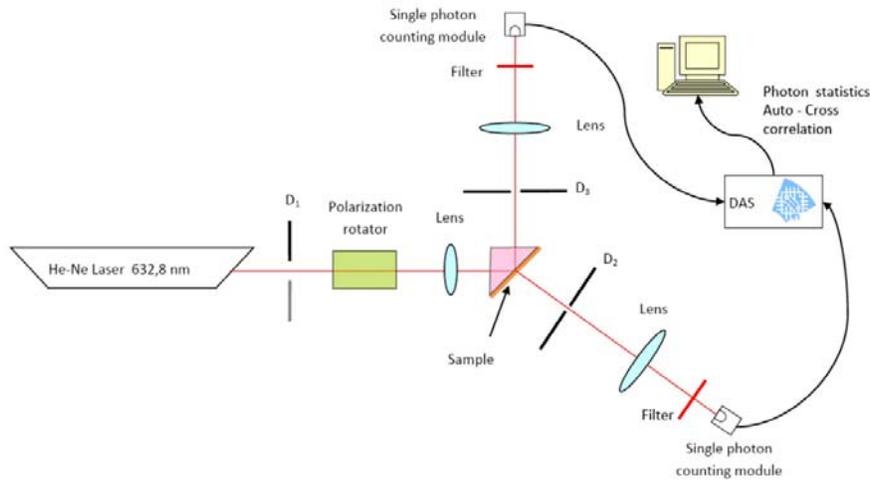

**Figure 4a.** Shows the outline of the first experimental setup, where the beam reflected from the metal layer (sample) and the light emitted from the opposite side of the layer by surface plasmons where measured with Perkin Elmer SPCM - AQR-14 type photon counting modules. The other details can be found in the main text.

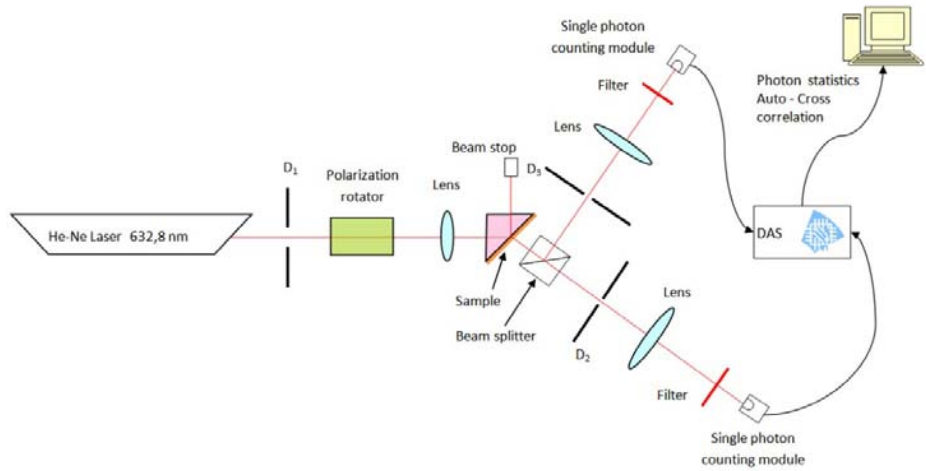

**Figure 4b.** Shows the second setup used by our experiments. The details of the arrangement are summarized in the text. The difference in comparison with the first setup shown in Fig. 4a is that now the light stemming from the SPO is split by the beam splitter, which has been placed between the target and the first detector, such that the second detector received the reflected light.

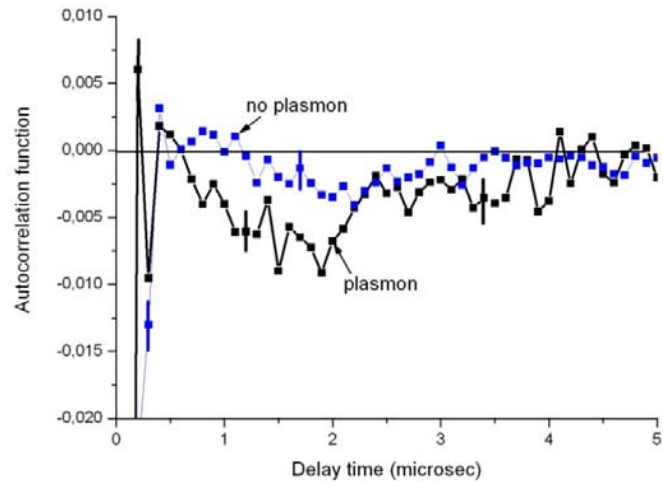

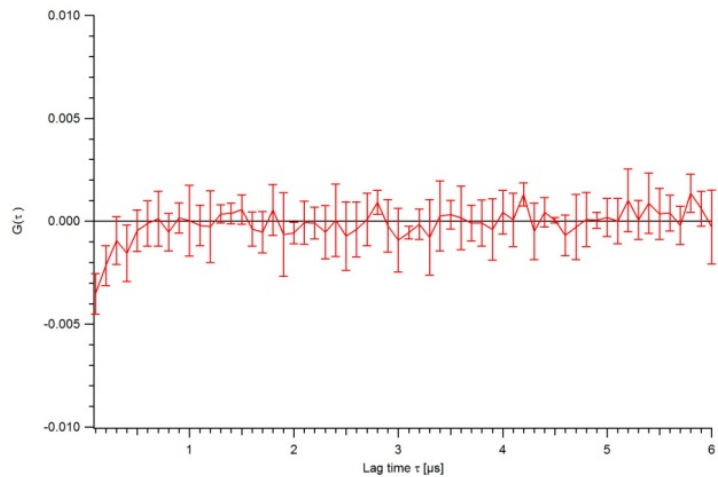

**Figure 5.** The upper figure (a) shows the auto-correlation function in the two arms (see Fig. 4a, setup 1). The note 'no plasmon' refers to that the reflected light has been sampled. We see that this signal is essentially poissonian within the accuracy of our measurements. The curve marked with 'plasmon' shows the dependence of the auto-correlation function of the light emitted by the SPOs. Here we see a slight (~0.5%) subpoissonian deviation from the shot-noise. In the lower figure (b) the measured 'cross-correlation' function is displayed. Here the anti-correlation is less pronounced (~0.3%). This is certainly due to the increased number of modes covered by the two detectors. In fact, in each cases the reduced covariance $R = \overline{N_1 N_2}/\overline{N_1}\,\overline{N_2} - 1$ has been measured, and this should be compared with the theoretical results shown in Fig. 3a-b.

The experimental results are in quite good qualitative agreement with the theoretical predictions, but the relatively modest resolution we had by now prevented us to make further and more detailed quantitative comparison.

## 4. Summary

In the present paper we have discussed the statistical properties of light emitted by suface plasmons (SPOs), genetared in a gold layer in the Kretschmann geometry. In Section 2 we have given a detailed analysis of the empirical parameters characterising the SPOs in the framework of classical electrodynamics, with a particular emphasis on the coherence properties. Besides, we have applied a new approach to describe the quantum statistical properties of the SPOs, as single-photon emitters. The most important parameters (like the Fano factor and the covariance of counting events in delayed or spatially separated detectors) of the photon distribution have been calculated on the basis of the obtained analytic formulas. The transition from anti-bunching to bunching has also been found and analysed. Though with the present analysis we have offered a reasonable interpretation of the measurements data, concerning the transverse coherence of the plasmon light, we have mostly relied on intuitive (but quantitative) reasonings. The theory should in particular, be developed further in this respect. In our first experimental results on correlation properties (auto-correlation and cross-correlation) of the SPO radiation, we have found evidence for the predicted nonclassical counting statistics. However, it is clear to us that refinements (like improving the time resolution, or applying the start-stop method for the data aquisition) of our technique are desirable for further progress.

**Acknowledgements**

This work has been supported by the Hungarian National Scientific Research Foundation OTKA, Grant No. K73728.

Varró S, Kroó N, Oszetzky D, Nagy A and Czitrovszky A, Hanbury Brown – Twiss type correlations with surface plasmon light. Published in J. Mod. Opt. 58, 2049-2059 (2011)[18] Mandel, L.; Wolf, E., *Phys. Rev.* **1961**, *124*, 1696-1702.

[19] Loudon, R.,. *Rep. Prog. Phys*. **1980**, *43*, 913-949.

[20] Mandel, L.; Wolf, E.: *Optical Coherence and Quantum Optics*. (Cambridge University Press, Cambridge, **1995**)

[21] Scully, M. O.; Zubairy, M. S.: *Quantum Optics*. (Cambridge University Press, Cambridge, **1997**)

[22] Loudon, R.: *The Quantum Theory of Light.* (Clarendon Press, Oxford, **2000**)

[23] Schleich, W. P.: *Quantum Optics in Phase Space.* (Wiley-VCH, Berlin, **2001**)

[24] S. Varró., *Physica Scripta* **T140**, 014038 (2010).

[25] J. Bergou and S. Varró, *J. Phys. A: Math. Gen.* **14**, l469-l482 (1981);

[26] Y. Ben-Aryeh and A. Mann, *Phys. Rev. Lett.* **54**, 1020-1024 (1985);

[27] W. Becker, K. Wódkiewicz and M. S. Zubairy, *Phys. Rev. A* **36**, 2167-2170 (1987).

[28] S. Varró, *New J. Phys.* **10**, 053028 (2008)(35pp);

[29] G. Bédard, Phys. Rev. **1966**, *151*, 1308-1039.

[30] H. Raether H, *Surface plasmons* (Berlin: Springer, 1988)

[31] A. V. Zayats, I. I. Smolyaninov , *J. Opt. A: Pure Appl. Opt.* **5**, S16 (2003)

[32] A. V. Zayats, I. I. Smolyaninov and A. A. Maradudin , *Physics Reports* **408,** 131 (2005)

[33] P. B. Johnson and R. W. Christy, *Phys. Rev. B* **6**, 4370 (1972)

[34] M. Born and E. Wolf, *Principles of optics*, Section 13.3 (London: Pergamon, 1959)

[35] Varró, S., Laser and Particle Beams, **2007**, *25*, 379-390 (2007)

[36] Goodman, J. W.: *Statistical Optics.* (John Wiley & Sons, Inc., New York, **1985**)

[37] Varró, S., *Fortschr. Phys*. **2011**, *59*, 296-324.

[38] Feller, W.: *An introduction to probability theory and its applications.* I. (John Wiley & Sons, Inc., New York, **1966**)

[39] Abramowitz, M.; Stegun, I. A., *Handbook of Mathematical Functions*. (Dover Publications, Inc., New York, **1974**)

**Figures and Captions to the Figures**

**Figure 1a**

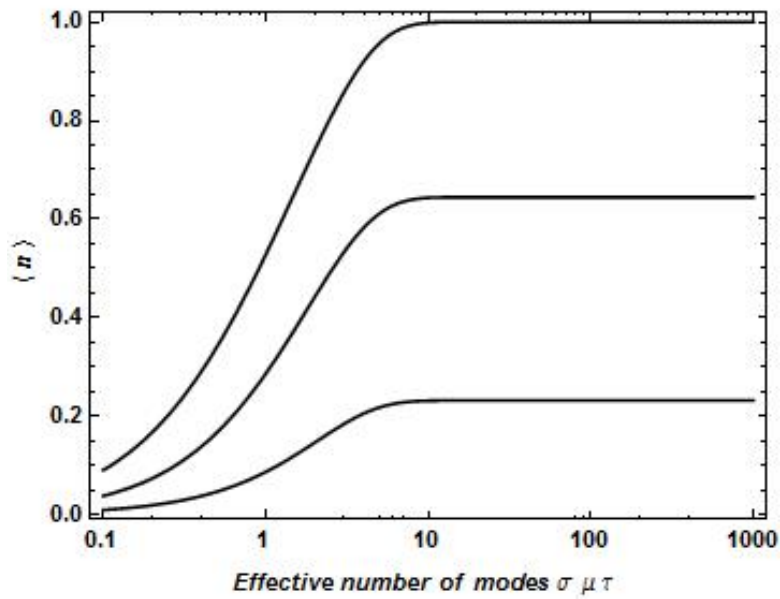

**Figure 1b**

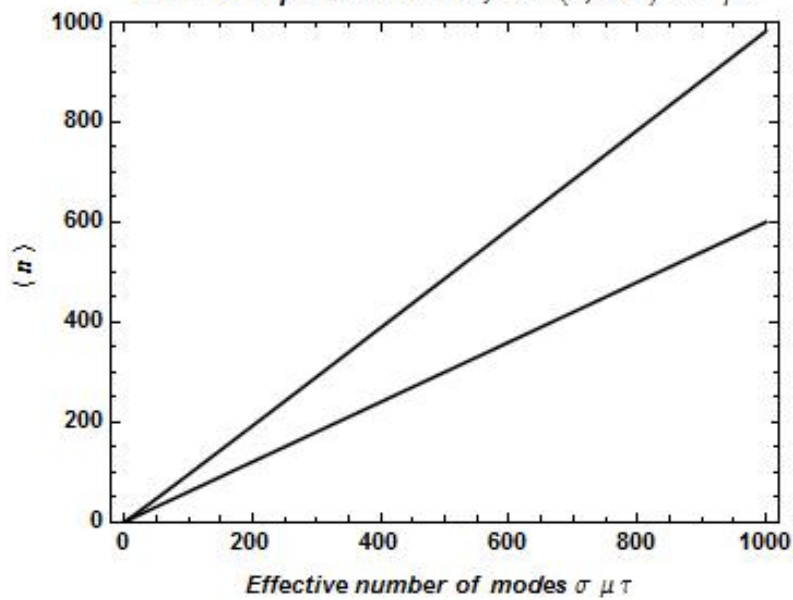

**Figure 2a**

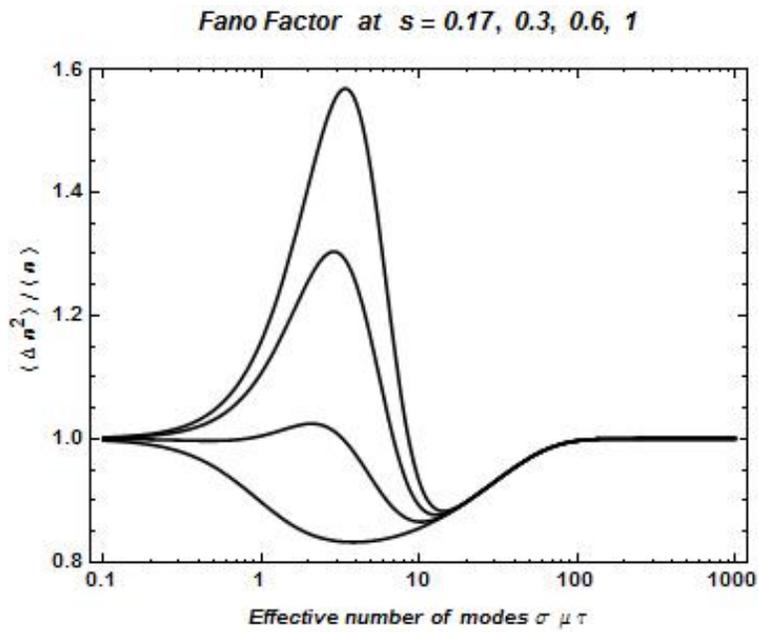

**Figure 2b**

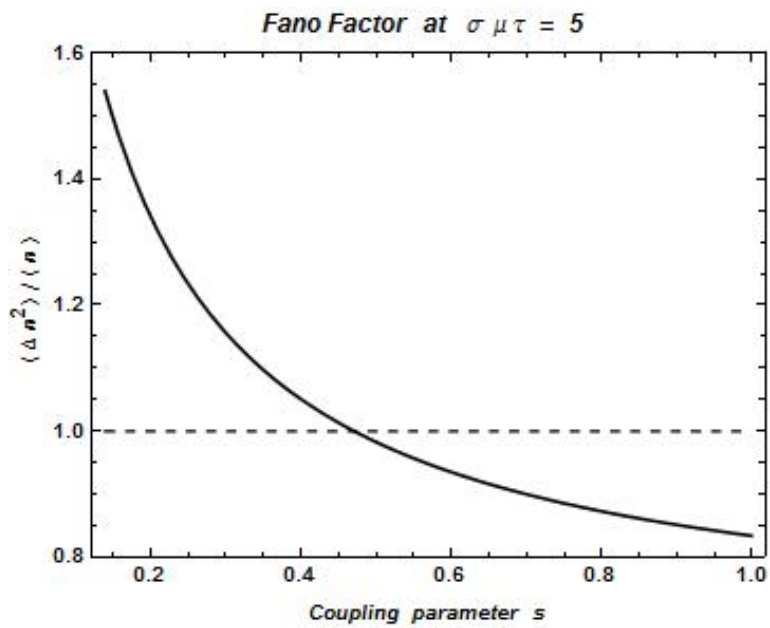

**Figure 3a**

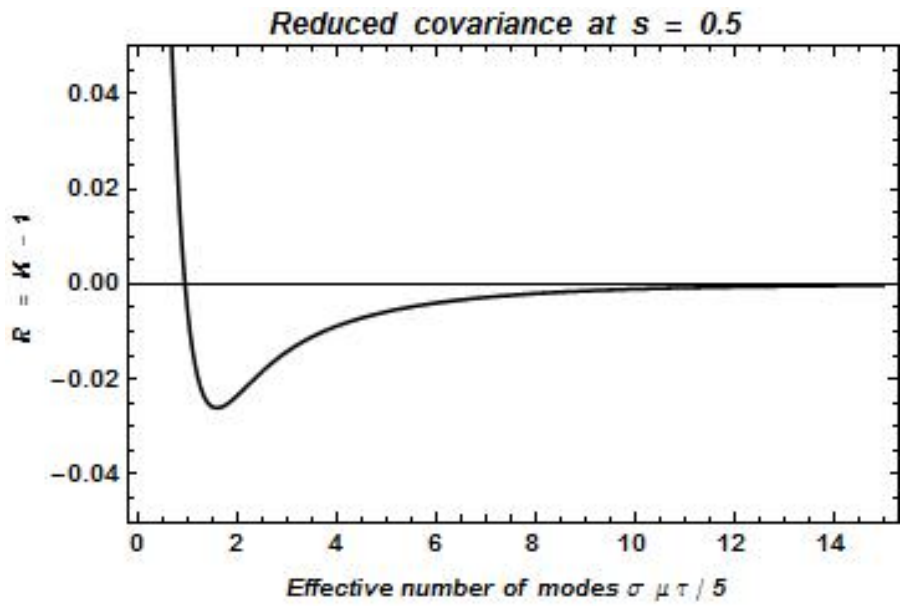

**Figure 3b**

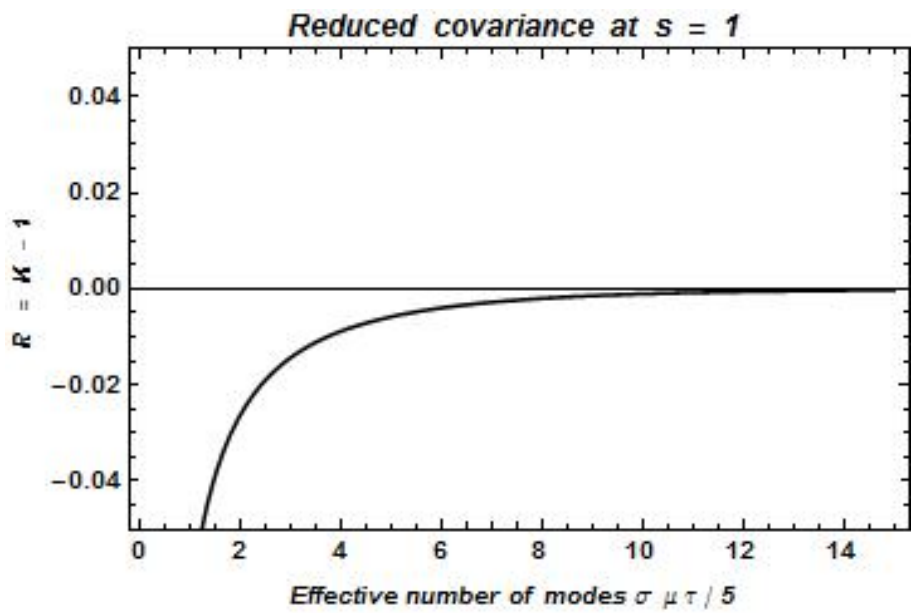

**Figure 4a**

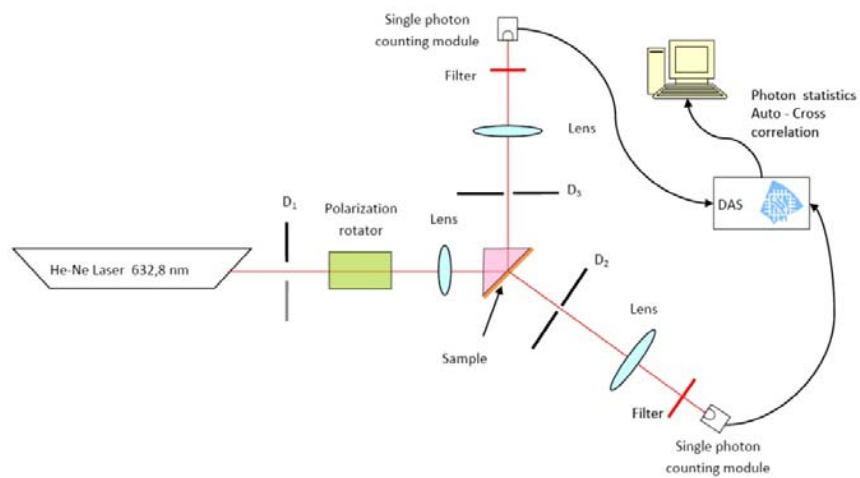

**Figure 4b**

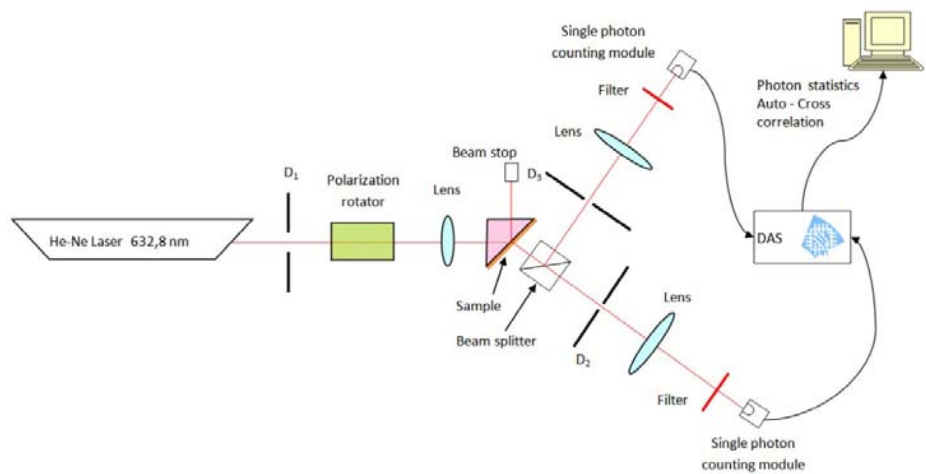

**<u>Figure 5a</u>**

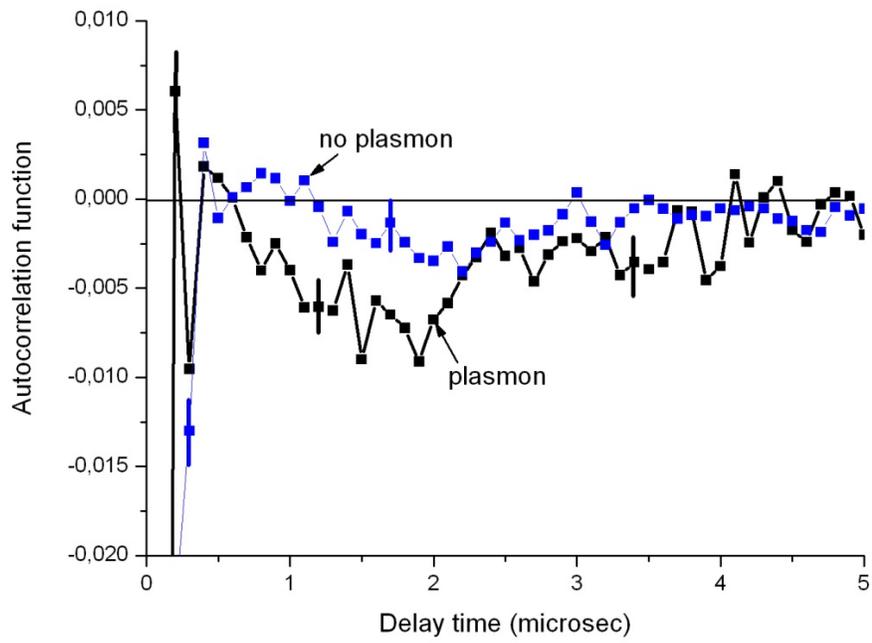

**<u>Figure 5b</u>**

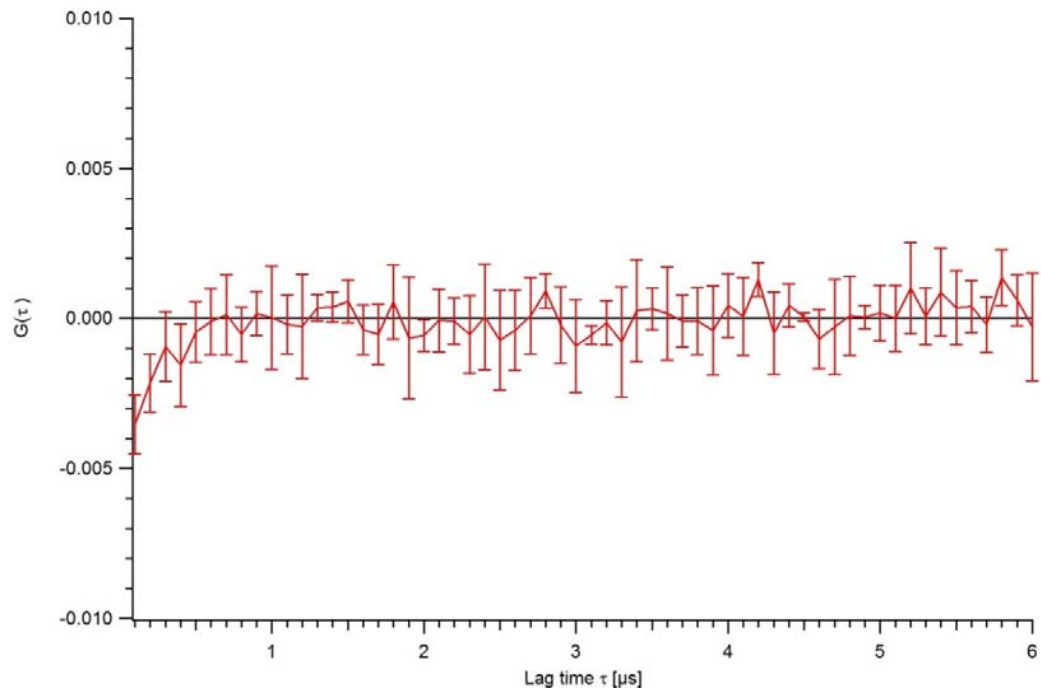

**Caption to Figures 1a-b**

**Figure 1.** Left (a): shows in a log-linear plot the dependence of the mean occupation number on the dimensionless interaction time $\sigma\mu\tau$, for three values of the coupling parameter $s$. Here the parameter $\lambda$ has been kept fixed. Right (b): same as in (a), but now we have taken $\lambda$ being proportional with interaction time: $\lambda = \eta \times \sigma\mu\tau$. In this case $\eta$ can be considered as a quantum efficiency (for which we have taken the values 1 and 0.6. Since the figure shows exact linear dependence, the mean number of electron counts is simply proportional with $\langle n \rangle$ in this case. This proportionality is not affected by the value of the coupling parameter $s$.

**Caption to Figures 2a-b**

**Figure 2.** Left (a): shows in a log-linear plot the Fano factor $F = \langle \Delta n^2 \rangle / \langle n \rangle$ on the dimensionless interaction time $\sigma\mu\tau$, for four values of the coupling parameter $s$. The uppermost curve corresponds to the lowest value $s = 0.17$, and lowermost one to the value $s = 1$. Right (b): the transition from bunching (super poissonian statistics) to antibunching (subpoissonian statistics) is illustrated at a particular value of the dimensionless combination $\sigma\mu\tau = 5$. This curve indicate that for very small coupling parameters, the correlation corresponds to more and more disordered (chaotic) distributions, formally $F \to \infty$ as $s \to 0$.

**Caption to Figures 3a-b**

**Figure 3.** Shows the reduced covariance $R = K - 1 = \overline{N_1 N_2}/\overline{N}_1\,\overline{N}_2 - 1$ calculated on the basis of Eqs. (5.a-b). This quantity, like the Fano factor plotted in Figs. 2a-b, is very sensitive to the value of the coupling constant. On the left figure we see a transition from anti-bunching to bunching as the scaled delay $\sigma\mu\tau$ decreases. On the right figure, since the coupling parameter is already large enough ( $s = 1$ ), no such transition appears, thus the anti-correlation characteristic in the single-photon regime dominates throughout.

**Caption to Figures 4a-b**

**Figure 4a.** Shows the outline of the first experimental setup, where the beam reflected from the metal layer (sample) and the light emitted from the opposite side of the layer by surface plasmons where measured with Perkin Elmer SPCM - AQR-14 type photon counting modules. The other details can be found in the main text.

**Figure 4b.** Shows the second setup used by our experiments. The details of the arrangement are summarized in the text. The difference in comparison with the first setup shown in Fig. 4a is that now the light stemming from the SPO is split by the beam splitter, which has been placed between the target and the first detector, such that the second detector received the reflected light.

**Caption to Figures 5a-b**

**Figure 5.** The upper figure (a) shows the auto-correlation function in the two arms (see Fig. 4a, setup 1). The note 'no plasmon' refers to that the reflected light has been sampled. We see that this signal is essentially poissonian within the accuracy of our measurements. The curve marked with 'plasmon' shows the dependence of the auto-correlation function of the light emitted by the SPOs. Here we see a slight (~0.5%) subpoissonian deviation from the shot-noise. In the lower figure (b) the measured 'cross-correlation' function is displayed. Here the anti-correlation is less pronounced (~0.3%). This is certainly due to the increased number of modes covered by the two detectors. In fact, in each cases the reduced covariance $R = \overline{N_1 N_2}/\overline{N}_1 \overline{N}_2 - 1$ has been measured, and this should be compared with the theoretical results shown in Fig. 3a-b.